\documentclass[9pt,twocolumn,twoside]{pnas-new}
\templatetype{pnasresearcharticle}
\usepackage{graphicx,amsmath,eqnarray}
\usepackage[outdir=./]{epstopdf}
\usepackage{dcolumn} 
\usepackage{bm} 
\usepackage{color,stmaryrd}

\def\S{\mathbf{S}}

\usepackage{xr}
\makeatletter
\newcommand*{\addFileDependency}[1]{
  \typeout{(#1)}
  \@addtofilelist{#1}
  \IfFileExists{#1}{}{\typeout{No file #1.}}
}
\makeatother

\newcommand{\eee}[1]{\mathrm{e}^{\normalsize #1 }}
\newcommand{\ii}{\mathrm{i}} 

\title{Spontaneous phase coordination and fluid pumping in model ciliary carpets}
\author[a]{Anup V Kanale}
\author[a]{Feng Ling}
\author[a,b,c]{Hanliang Guo}
\author[d,e]{Sebastian F\"urthauer}
\author[a,1]{Eva Kanso}
\affil[a]{Aerospace and Mechanical Engineering, University of Southern California, Los Angeles, CA 90089, USA.}
\affil[b]{Department of Mathematics, University of Michigan, Ann Arbor, MI 48109, USA.}
\affil[c]{Department of Mathematics and Computer Science, Ohio Wesleyan University, Delaware, OH 43015, USA}
\affil[d]{Center for Computational Biology, Flatiron Institute, New York, NY 10010, USA.}
\affil[e]{Institute for Applied Physics, TU Wien, 1050 Wien, Austria.}

\date{\today}

\leadauthor{Kanale, Ling} 

\significancestatement{
The human lungs, brains, and reproductive tracts contain ciliated tissues that are specialized in pumping fluids. These tissues generate flows by the collective activity of hundreds of thousands of cilia, where each cilium beats similarly to its neighbors, but with a different phase depending on its position, creating a metachronal wave pattern. Despite its importance for tissue-level function, the emergence of stable ciliary waves is not well understood. Here, we develop discrete and continuum theories for (infinitely)-large arrays of hydrodynamically coupled cilia, and identify minimal requirements for the emergence of stable collective coordination and fluid pumping. Our framework opens up the prospect to evaluate and design ciliated tissues that break symmetry and pump fluids in biological and engineered systems.
}
\authorcontributions{}
\authordeclaration{There is no conflict of interest to declare.}
\equalauthors{AVK and FL contributed equally to this work.}
\correspondingauthor{\textsuperscript{1}To whom correspondence should be addressed. E-mail: kanso@usc.edu}
\begin{abstract}
Ciliated tissues such as in the mammalian lungs, brains, and reproductive tracts, are specialized to pump fluid. They generate flows by the collective activity of hundreds of thousands of individual cilia that beat in a striking metachronal wave pattern. Despite progress in analyzing cilia coordination, a general theory that links coordination and fluid pumping in the limit of large arrays of cilia remains lacking. Here, we conduct in-silico experiments with thousands of hydrodynamically-interacting cilia, and we develop a continuum theory in the limit of infinitely-many independently beating cilia by combining tools from active matter and classical Stokes flow. We find, in both simulations and theory, that isotropic and synchronized ciliary states are unstable. Traveling waves emerge regardless of initial conditions, but the characteristics of the wave and net flows depend on cilia and tissue properties. That is, metachronal phase coordination is a stable global attractor in large ciliary carpets, even under finite perturbations to cilia and tissue properties. These results support the notion that functional specificity of ciliated tissues is interlaced with the tissue architecture and cilia beat kinematics and open up the prospect of establishing structure-to-function maps from cilium-level beat to tissue-level coordination and fluid pumping.
\end{abstract}
\dates{This manuscript was compiled on \today}

\begin{document}

\maketitle

\thispagestyle{firststyle}
\ifthenelse{\boolean{shortarticle}}{\ifthenelse{\boolean{singlecolumn}}{\abscontentformatted}{\abscontent}}{}

\dropcap{M}otile cilia are basic micro-actuators in cell biology and bio-inspired systems~\cite{Gilpin2020} that beat cyclically~\cite{Man2019} to transport fluid across the cell surface.
In aquatic species, cilia can be found along external and internal epithelial surfaces in solitary, pairwise, or lattice configurations, where they have a broad array of functions, from locomotion~\cite{Wan2016} to food capture~\cite{Ding2015,wan2020reorganization} and acquisition of microbial partners~\cite{Nawroth2017}. 
When animals invaded terrestrial habitats, cilia became restricted to internal epithelial surfaces to reduce water loss across ciliary membranes, rendering them difficult subjects for direct study.
In humans, they serve important biological functions, including mucus clearance in the respiratory system~\cite{Ramirez2020}, left–right asymmetry determination during embryonic development~\cite{Hirokawa2009}, cell transport in the reproductive tracts~\cite{Raidt2015}, and cerebrospinal fluid circulation in the brain ventricles~\cite{Faubel2016,Pellicciotta2021}.

To pump fluid at the micron scale where viscosity is dominant~\cite{Purcell1977}, individual cilia beat in a non-reversible manner and, in multi-ciliated cells and tissues, neighboring cilia coordinate their phase to beat in metachronal waves~\cite{Michelin2010,Osterman2011,Elgeti2013,Ding2014,Guo2021b}.
Metachronal coordination and microscale flows were observed in microdissected \textit{ex-vivo} epithelia~\cite{Faubel2016, Nawroth2017,Ramirez2020} and  \textit{in-vitro} engineered tissues~\cite{Pellicciotta2021}. However,
probing the emergence and robustness of cilia coordination in tissues, and its effect on fluid transport, remains experimentally challenging.

Microorganisms provide an accessible way to investigate cilia coordination, but functional constraints on coordination differ between organisms that use cilia for locomotion and tissues that use cilia for specialized fluid pumping functions.
Specialization and efficient pumping in tissues can be compromised if multiple states of cilia coordination co-exist.
In contrast, to achieve and transition between swimming and turning in cilia-driven locomotion, multiple coordination states are needed~\cite{Polin2009, Wan2016,cortese2021control,wan2021flagella}. In-phase, antiphase, and non-trivial phase lags were observed experimentally in pairs of cilia isolated from the somatic cells of \textit{Volvox carteri}~\cite{Brumley2014} and \textit{Chlamydomonas reinhardtii}~\cite{Wan2016}, and coupled only via the fluid medium. These coordination states, and the instabilities that lead to transitions between them, were captured quantitatively \textit{in-silico}~\cite{Man2020}.
In addition to interactions via the fluid medium, 
cilia coordination in unicellular organisms seem to be communicated predominantly
by elastic coupling through the cytoskeleton~\cite{Geyer2013,Wan2016, Klindt2017, Liu2018, Guo2021}. However, the importance of substrate-based interactions in tissues, besides providing and maintaining geometric and structural integrity, remains under debate, with recent experimental evidence  suggesting that fluid coupling dominates cilia coordination in tissues~\cite{Pellicciotta2021}.

Are hydrodynamic interactions sufficient to guarantee functional specificity and coherent fluid pumping in ciliated tissues? The answer requires an investigation of the coordination states that emerge at the tissue level and how they drive flows. Mathematical models and numerical simulations offer exciting prospects for exploring these issues, however, most existing models address each aspect separately: cilia coordination ~\cite{Guo2018,Man2020,Guo2021,Meng2021,Solovev2020,Chakrabarti2021} or cilia-generated flows~\cite{Ding2014, Ramirez2020}. Meng \textit{et al.} analyzed, in a coarse-grained model, the stability of metachronal waves as function of the dynamical and geometric characteristics of the individual cilium~\cite{Meng2021},
and, Solovev \textit{et al.} showed, using a similar analysis, that metachronal waves are globally stable~\cite{Solovev2020}.
Ramirez \textit{et al.} measured ciliary arrangement from direct images of \textit{ex-vivo} mammalian airways, and reported heterogeneity in cilia organization across the trachea but coherent fluid transport~\cite{Ramirez2020}.
The results in \cite{Meng2021} and \cite{Ramirez2020} are important in complementary ways: \cite{Meng2021} provided tools to analyze how collective properties depend on individual beat patterns, while \cite{Ramirez2020} provided tools to map collective properties of the ciliated tissues to fluid transport function. 
The combined effect of how cilia in large scale ciliary carpets self-organize into emergent collective states and how these states drive flows is less understood~\cite{Osterman2011, Elgeti2013}. Here, we investigate coordination and pumping in discrete and coarse-grained models of ciliary carpets. Our findings challenge the basic conception in \cite{Meng2021} that wave stability is the only factor that matters for tissue-level function and expand the observations in~\cite{Meng2021,Ramirez2020}.

Our models serve as a first step towards establishing maps from single-cilium kinematics to tissue-level coordination and fluid pumping function, and could help inform our understanding of the function, and dysfunction, of ciliated tissues in major mammalian organs. They could also help translate cilia-inspired design principles to engineered micro-pumps~\cite{Khaderi2012,Hanasoge2018, Shields2010,Milana2020}; a potential that is particularly compelling in light of the rapid developments in tissue engineering and organ-on-chip~\cite{nawroth2021breathing,nawroth2021modular} technologies.

\section*{Results}

\begin{figure*}[!t]
	\centering
	\includegraphics[scale=1.1]{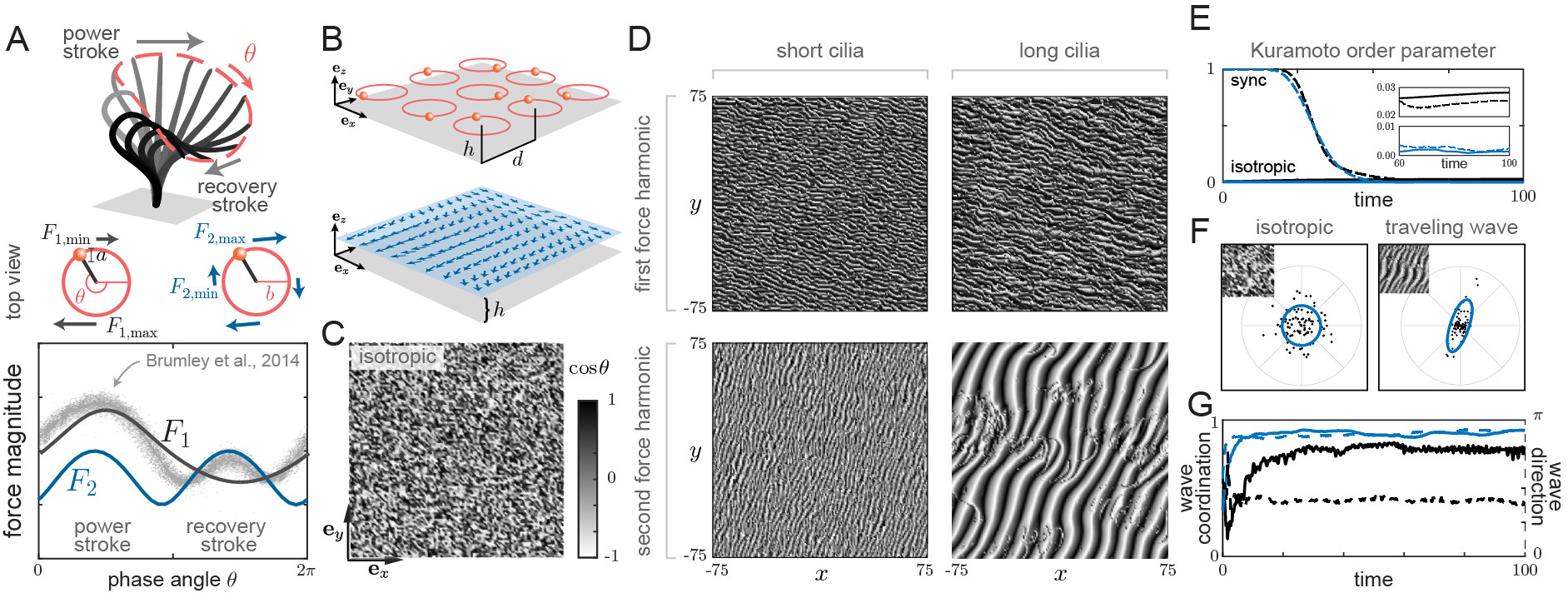}
	\caption{\textbf{Ciliary carpets:}
	\textbf{A.} Cilia beats usually has a stronger power stroke and a recovery stroke, and typically have a preferred beating axis in which it travels a longer distance.
	These two features can be individually characterized by the first two harmonic decomposition of a periodic forcing produced by a rotating force monopole situated at a fixed height away from a no-slip wall.
	Specifically, the first force harmonic $F_1$ captures the force magnitude variation due to power and recovery stroke, while the second force harmonic $F_2$ can represent the force variation due to an elliptical beat.
	These observations can be confirmed by matching the harmonic force magnitudes with those deduced from real cilia beats as a function of the phase angle $\theta$.
	Both the 3D beat and the force profile are traced from \cite{Brumley2014}.
    \textbf{B.} A doubly-periodic array of cilia on a square lattice in the $(x,y)$ plane above a wall is represented in the continuum limit by a force discontinuity layer shown in blue.
	\textbf{C.} Discrete simulation of the field of rotors can be succinctly represented by the cosine of their phase. Here an isotropic state was used to initialize simulations of 151$\times$151 cilia.
	\textbf{D.} Numerical simulations showing the emergence of traveling metachronal waves. Snapshots taken at $t=100$ for $h/d=0.5$ (left) and $h/d=1.5$ (right). 
	\textbf{E.} Time evolution of the Kuramoto polar order parameter $P$ shows no clear distinction between isotropic and traveling wave patterns. First force harmonic is shown in black and second harmonic in blue. 
	\textbf{F.} We define $P(\alpha)$ for $\alpha \in [0,2\pi]$ and plot $(\alpha, P(\alpha))$ in polar coordinates at two snapshots taken from the isotropic state (panel B) and traveling wave pattern (bottom left of panel C). Data points are shown in black and elliptic best-fit in blue. 
	\textbf{G.} Time evolution of wave coordination (Kuramoto ellipse eccentricity) and wave direction (parallel to ellipse minor axis) clearly shows the growth of the traveling wave instability (see SI Movie). First and second force harmonics are shown in black and blue respectively, and solid and dashed lines represent wave coordination and direction respectively.
	In panel D-G, simulations are performed for $100$ time units using $dt=0.1$,  $a=0.05$, $b=0.2$, $d=1$. First force harmonic: $F_1 = 1 + 0.5 \cos\theta + 0.5 \sin\theta$, Second force harmonic: $F_2 = 1 + 0.5 \cos2\theta + 0.5 \sin2\theta$.
    }
	\label{fig:setup}
\end{figure*}

To investigate the emergent coordination and fluid pumping in tens of thousands of hydrodynamically-coupled cilia,
we develop discrete and coarse-grained models where individual cilia are represented as nonlinear phase oscillators
in a viscous fluid medium of viscosity $\eta$
(Fig. \ref{fig:setup}A). 
We use our models to investigate three questions that are fundamental for understanding the physics of ciliated tissues. First, considering cilia whose individual beat patterns break no symmetry and produce no net flux at the single-cilium level, are hydrodynamic interactions in ciliary carpets  sufficient to create coordination patterns that break symmetry and pump fluid?
If so, how does the net flux created by this emergent coordination depend on the properties of the single-cilium?
Lastly, how do variations in the tissue-level heterogeneity~\cite{Ramirez2020,Boselli2021}
affect both cilia coordination and net flux?

\paragraph{Modeling individual cilia as nonlinear phase oscillators.} 
We represent a beating cilium by a spherical bead of radius $a$ moving on a circular trajectory of radius $b$ in an $(x,y)$-plane located at a height $z=h$ above a bounding wall at $z=0$ (Fig.~\ref{fig:setup}).
In this representation, phase $\theta$ is the only dynamical variable; properties of the cilium beating pattern are subsumed to a phase-dependent active force $\mathbf{F}(\theta)$ that drives the bead's motion. 
The fluid velocity $\mathbf{u}(x,y,z)$ is governed by the incompressible Stokes equations and obtained from the Blake-Oseen solution associated with the force monopole $\mathbf{F}(\theta)$ in the 3D half-space ($z\geq 0$).
Common, including experimentally-derived, cilia beat patterns, can be accounted for by fitting $\mathbf{F}$ such that the flow field $\mathbf{u}$ matches that of the beating cilium \cite{Brumley2014}.

We restrict our analysis to when the active force $\mathbf{F} = F(\theta)\mathbf{t}$ acts in the tangent direction $\mathbf{t}$ to the circular trajectory (Fig.~\ref{fig:setup}A). 
Generically,  $F(\theta)$ is a strictly-positive, nonlinear function that can be expressed in terms of force harmonics, 
$F(\theta) = F_o\left[ 1 + \sum_{m=1}^\infty \alpha_m \cos m\theta + \beta_m \sin m \theta \right]$;  $F_o$ is a constant force and $\alpha_m, \beta_m$ are non-dimensional coefficients. The first harmonic ($m=1$) defines a distinguishable direction in the $(x,y)$-plane from minimum to maximum forcing, akin to the power stroke direction in beating cilia; the second harmonic ($m=2$) describes a forcing profile with two antipodal maxima orthogonal to two antipodal minima, thus capturing the elliptic characteristic of a beat cycle with no asymmetry (SI, $\S$2, Fig.~S1).

The tangential active force ensures no fluid pumping at the single cilium level. The net flux $q = \|(1/2\pi)(h/\pi\eta)\int_0^{2\pi} \mathbf{F}d\theta \|$ is identically zero. 
Allowing a non-tangential force component or tilting the ciliary trajectory with respect to the bounding wall can break symmetry and prompt fluid pumping~\cite{Smith2008,Meng2021}. Here, we confine our analysis to rotors that do not break the Stokes flow symmetry to distill whether hydrodynamic interactions in ciliary carpets can lead to symmetry-breaking and fluid pumping.

\paragraph{Modeling ciliary carpets.} We consider a two-dimensional (2D) discrete array of cilia, where each cilium is centered at $\mathbf{x}_j$ ($j=1,2,\ldots$) in a doubly-periodic square lattice of fundamental domain size $L\times L$ and distance $d$ between neighboring cilia (Fig.~\ref{fig:setup}A). Let $N$ be the total number of cilia in the fundamental domain and $\rho_c = N/L^2$ the cilia density.

The instantaneous position of the $j^{th}$ cilium is determined by its phase $\theta_j$, and, in vector form, by $\mathbf{r}_j= \mathbf{x}_j + b\mathbf{n}_j$, where $\mathbf{n}_j=(\cos\theta_j,\sin\theta_j,0)$ 
is the normal unit vector in the ciliary plane. 
Each cilium  is driven independently by an active tangent force $\mathbf{F}_j=F(\theta_j)\mathbf{t}_j$.
The fluid velocity $\mathbf{u}(x,y,z)$ 
driven by the collective beating of all cilia is given by a conditionally-convergent, doubly-infinite sum of the Blake-Oseen tensor (SI, $\S$2, Figs.~S2 and S7). For notational convenience, we introduce the 2D velocity 
$\mathbf{v}(x,y)$, which is the projection of $\mathbf{u}(x,y, z=h)$ onto the ciliary plane. 

Balance of forces on cilium $j$ in the ciliary plane dictates that $-\zeta \left( b\dot{\theta}_j\mathbf{t}_j - \mathbf{v}(\mathbf{r}_j)\right)  + \mathbf{F}_j + \mathbf{N}_j= 0,$ where the first term denotes the drag force (with constant drag coefficient $\zeta = 6\pi \eta a$) and accounts for hydrodynamic coupling among cilia; the last term is a constraint force  that guarantees the bead remains on the desired circular trajectory. 
Projecting the force balance on the normal and tangential directions, respectively, leads to an expression for the constraint force, $\mathbf{N}_j = -\zeta(\mathbf{n}_j\cdot \mathbf{v}(\mathbf{r}_j))\mathbf{n}_j$, and a set of coupled differential equations that govern the time-evolution of the cilia phase,
\begin{equation}
    \dot{\theta}_j = \Omega_j + \dfrac{1}{b}\mathbf{t}_j\cdot \mathbf{v}(\mathbf{r}_j),
    \label{eq:eom_theta}
\end{equation}
with intrinsic phase-dependent angular speed $\Omega_j = F(\theta_j)/\zeta b$.

\paragraph{Collective states that do not pump fluid} We look for time-periodic, spatially-uniform states in which the collective dynamics is time reversible, or equivalently,~\eqref{eq:eom_theta} is invariant under the transformation  $\theta \to -\theta$. 
Two collective states exist: synchronized ($\theta_i=\theta$) and isotropic (Fig.~\ref{fig:setup}C).
When all cilia beat in synchrony, we get
$\theta_j(t) = \theta^\ast(t)$, $\mathbf{t}_j = \mathbf{t}^\ast$,  and $\mathbf{v}(\mathbf{r}_j) = v^\ast\mathbf{t}^\ast$ for all $j$. This is a periodic solution of~\eqref{eq:eom_theta}, for which the dynamics is time-reversible and no pumping ensues. 

To define an isotropic state for which the fluid velocity is identically zero, i.e. $\mathbf{v}=0$, we first introduce a  nonlinear coordinate transformation ${d\phi}/{d\theta} = {\Omega_o}/{\Omega(\theta)}$ from $\theta$ to a new phase $\phi$ in which the intrinsic angular speed $\Omega_o=2\pi/T_o$ is constant and $T_o = \int_0^{2\pi} d\tilde\theta/\Omega(\tilde\theta)$ is the oscillation period of isolated cilia (SI,$\S$4 Fig.~S11).  We rewrite~\eqref{eq:eom_theta} in terms of $\phi$,
\begin{equation}
    \dot{\phi}_j = \Omega_o +\dfrac{1}{b}\dfrac{\Omega_o}{\Omega_j}\mathbf{t}_j\cdot \mathbf{v}(\mathbf{r}_j).
    \label{eq:eom_discrete_phi}
\end{equation}
We define the isotropic state to be the state where all $\phi_j$ are drawn from a uniform distribution function $\text{Unif}\,[0,2\pi)$. In the limit of an infinitely dense ciliary carpet,  cilia-induced forces cancel everywhere in the fluid domain, resulting in a fluid velocity $\mathbf{v}$ that is identically-zero. The isotropic state is thus a periodic solution of~\eqref{eq:eom_theta} that does not pump fluid. In systems with finite numbers of cilia, the isotropic state can only be approximately realized.

\paragraph{Spontaneous phase coordination.} We probe the stability of the synchronized and isotropic states numerically, by solving~\eqref{eq:eom_theta} for $151\times 151$ cilia (a total of $N=22,801$ hydrodynamically-coupled cilia) using an in-house algorithm (see SI).
We find that, starting from initially synchronized and isotropic states, as time evolves, the cilia spontaneously coordinate into traveling wave patterns. The emergent patterns are independent of initial conditions, but depend on the force profile $F(\theta)$ and distance $h$ from the wall; Fig.~\ref{fig:setup}(D) shows the long-term travelling wave patterns corresponding to the first and second force harmonics at two values of $h$. 
Both the isotropic and the synchronized states of the system are unstable, giving rise to symplectic traveling waves.
A direct inspection of these emergent waves suggests that the coordination is more spatially coherent when subject to second force harmonic, consistent with \cite{Meng2021}. 
The role of the first force harmonic in emergent coordination was not discussed in \cite{Meng2021}.

\paragraph{Quantifying emergent wave patterns.}
In an effort to quantify the emergent order, we introduce the moment fields describing phase coordination~\cite{Fuerthauer2013,chakrabarti2022self},
\begin{equation}
\label{eq:Yn}
    {Y}_n(\mathbf{x},t) = \frac{1}{\rho_c} \sum_j^{N} \eee{\ii n \phi_j}\delta(\mathbf{x} - \mathbf{x}_j) ,
\end{equation} 
where ${Y}_1$ is the Kuramoto order field, $Y_2$ the nematic order field, and $\delta(\cdot)$ the Dirac delta function.
The average of $Y_1$ over the 2D fundamental domain leads to the Kuramoto order parameter $P = \left| \langle  Y_{1} \rangle_{\textrm{2D}} \right|$; 
values of $P$ near zero indicate phase disorder while values near one correspond to phase synchrony~\cite{Kuramoto2003}. 
In Fig~\ref{fig:setup}E, we plot $P$ versus time for the synchronized and isotropic initial state: $P$ converges to a steady-state value that is independent of initial conditions, indicating that these emergent waves are global attractors of the collective dynamics. 
However, $P$ does not capture the change in spatial patterns that arise as the system evolves from isotropic to  traveling wave patterns.

To characterize the emergent wave patterns, we compute a directional Kuramoto order parameter $P_\alpha = \left|\langle Y_1\rangle_{\textrm{1D}} \right|$
by integrating $Y_1$ along a straight line at a direction $\alpha$ from the $x$-axis.
Intuitively, $P_\alpha$ measures the average phase order along the direction $\alpha$. Mathematically, it is the absolute value of the Radon transform of $Y_1(\mathbf{x})$ restricted to lines that go through the origin. We vary $\alpha \in [0,2\pi)$ and compute $(\alpha, P_\alpha)$ (black dots in Fig.~\ref{fig:setup}F). The distribution of data points $(\alpha, P_\alpha)$ is roughly circular in the isotropic state and elliptic in the traveling wave state (Fig.~\ref{fig:setup}F).
We use principal component analysis to fit $(\alpha, P_\alpha)$ to an ellipse, that we call the \textit{Kuramoto ellipse} (SI, $\S$3, Figs.~S8 and S9).
Starting from the isotropic state, the time evolution of the eccentricity and angle of the Kuramoto ellipse are good indicators of the emergent traveling wave pattern: its eccentricity indicates wave coordination and its angle is orthogonal to the wave traveling direction (Fig.~\ref{fig:setup}G).

\paragraph{Emergent waves and fluid pumping depend on cilium beating pattern.} Our goal is to examine how these collective emergent waves break time-reversal symmetry and pump fluid. We first analyze how the active force $F(\theta)$ affects emergent coordination.

\begin{figure*}[!ht]
	\centering
	\includegraphics[scale=1]{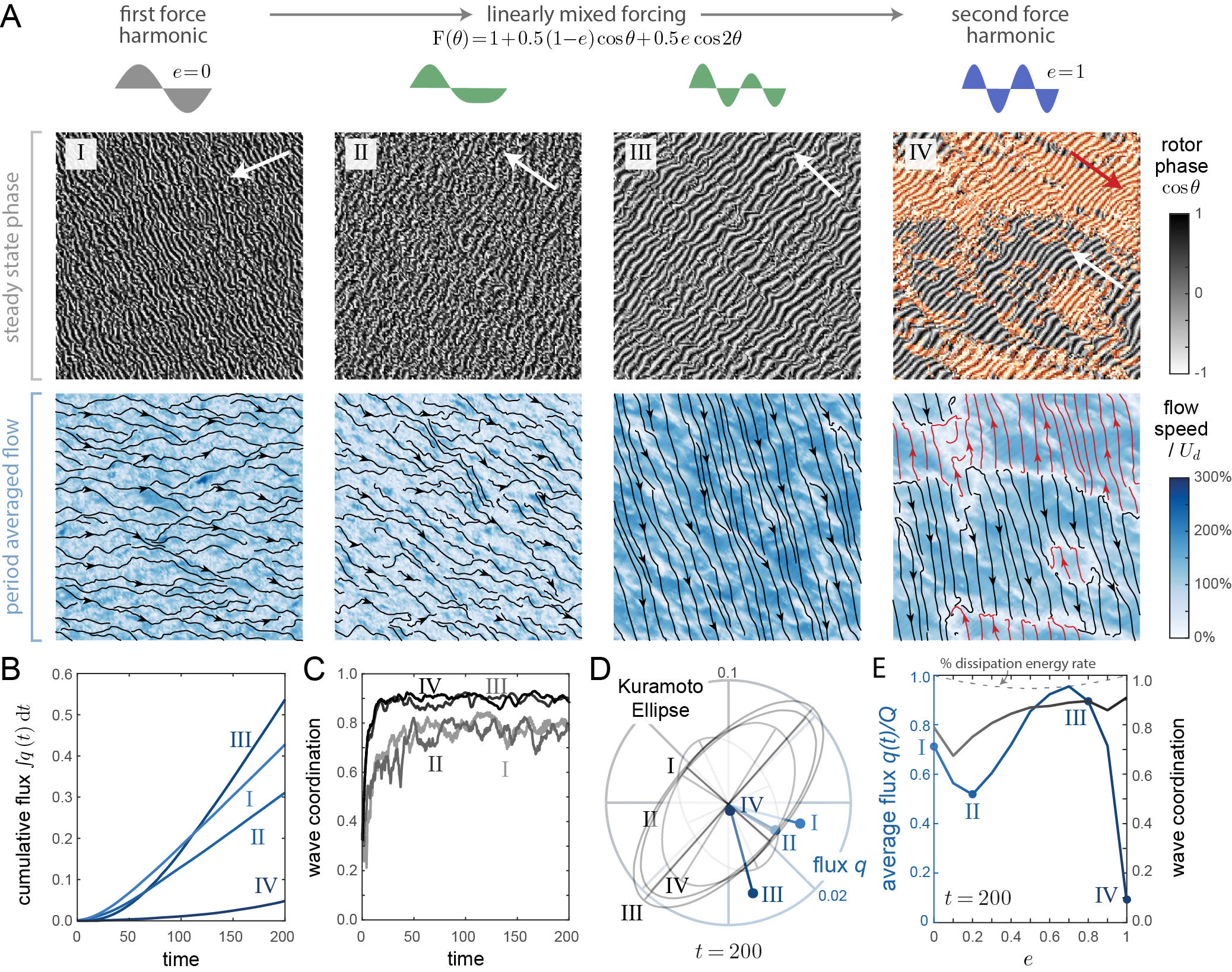}
	\caption{\textbf{Mixing first and second force harmonic produces maximum pumping.} \textbf{A.} Snapshot of a $151\times151$ ciliary lattice in steady state for (I) first force harmonic ($e=0$), (II) mixed forcing with larger first harmonics ($e=0.2$), (III) mixed with larger second harmonics ($e=0.8$), and (IV) pure second force harmonic ($e=1$). Tinted portion in (IV) indicates metachronal wave traveling in the opposite direction to the non-tinted portion.
	Here white arrows indicate the dominant wave direction and the red arrow shows the opposing direction for IV.
	On the second row, we show streamlines of the period averaged induced velocity field of the full system. We observe that while III produces the cleanest directed flow at highest speed, IV shows fluid being pumped in opposing directions.
	\textbf{B.} Cumulative flux magnitude $\int q\,\mathrm dt$ over 200 units of time shows that while III took a longer time to reach a high flux, it dominates over other presented cases eventually.
	\textbf{C.} Wave coordination as measured by eccentricity of Kuramoto ellipse. The higher the weight of second force harmonic is, the more likely that the wave becomes highly coherent.
	\textbf{D.} Kuramoto ellipse and net flux vector plotted in polar coordinates. Note that both I and III have the flux point roughly 30$^\circ$ clockwise of the respective minor axis, suggesting that a similar type of phase coordination is reached in these two cases.
	\textbf{E.} As the weight for second force harmonic increases, both the average flux and wave coordination first register a dip. This can be explained by the fact that the percent of energy put into the fluid (or rate of dissipation energy) is smaller when $e\neq0,1$; see SI. When $e$ reaches to about 3/4, we see that average flux is maximized. As $e$ continue to rise, the coordination improves while flux drops quickly due to the appearance of opposing traveling waves as shown in IV.
	} \label{fig:pumping}
\end{figure*}

The polarity introduced by the first force harmonic in the direction from minimum to maximum forcing sets a global direction for symmetry-breaking and fluid pumping.
Indeed, for $F(\theta) = F_o(1+\alpha_1\cos\theta + \beta_1\sin\theta)$, the emergent waves are spatially homogeneous and propagate in the same direction dictated by atan$(\beta_1/\alpha_1)$  for any choice of parameters $\alpha_1$ and $\beta_1$ and height $h$
(Figs.~\ref{fig:setup}D,~\ref{fig:pumping}A, case I, and Fig.~S10). 
For $F(\theta) = F_o(1+\alpha_2\cos2\theta + \beta_2\sin2\theta)$, the emergent waves propagate in two opposite directions (Fig.~\ref{fig:pumping}A, case IV), irrespective of parameter choice. That is, while both force harmonics break time-reversal symmetry locally, with more noticeable coordination in the second force harmonic, the latter does not lead to a global direction of emergent coordination and fluid pumping over the entire domain. This is because of the nematic symmetry of the second force harmonic $F(\theta+\pi) = F(\theta)$. 

To test these observations, we introduce a family of forcing $F(\theta) = 1 + 0.5(1-e) \cos(\theta) + e \cos(2\theta)$ that linearly morphs from first to second harmonic as $e$ varies from 0 to 1. In Fig.~\ref{fig:pumping}A-D, we show snapshots of the emergent coordination and fluid streamlines and speed for representative values of $e$.  We find that the first harmonic is essential in setting a global direction of phase coordination and fluid motion, whereas the second harmonic provides longer range of coordination. Importantly, while the wave coordination varies roughly monotonically as $e$ increases (Fig.~\ref{fig:pumping}C,E), with strongest phase coordination at $e=1$, the average flux $q$ depends nonlinearly on $e$ and is optimal for $e$ close but not equal to 1 (Fig.~\ref{fig:pumping}A, case III, and Fig.~\ref{fig:pumping}B,E).
At these optimal conditions, the second harmonic drives the instability, ensuring a strong metachronal coordination, while the first harmonic provides a weak perturbation that sets the global wave direction. 
At $e=1$, the first force harmonic makes zero contribution, metachronal waves propagate in opposite directions resulting in minimal net flux (Fig.~\ref{fig:pumping}A, case IV). 

Taken together, our results demonstrate that the first force harmonic, that emulates the effect of asymmetric beating, is essential for ensuring global coherence and spatial homogeneity of the wave pattern, whereas the second force harmonic, that represents the ellipticity of the cilia beating pattern, is more effective in driving the instability and achieving more coherent metachronal coordination.

\paragraph{Continuum model in the limit of infinite number of cilia} Our goal is to derive stability conditions and predict growth rates of the onset of instabilities that we observe in the isotropic and synchronized states and that lead to emergent coordination. A continuum theory was presented in~\cite{Meng2021}, providing important tools for assessing the linear stability of perfect metachronal waves. However, the theory in~\cite{Meng2021} lacks two important features: a method for evaluating a continuum fluid velocity field in the 3D half-space, and a mechanism for accounting for the onset of instabilities from non-coordinated states, such as the isotropic state. 
Here, we develop a continuum theory in the limit of dense cilia ($N\to\infty$), including an analytical derivation of the Stokes kernel associated with the ciliary layer following~\cite{Masoud2014, Gao2015}, that is applicable to coordinated and non-coordinated initial states. 

We start by modeling the ciliary carpet as a force density layer (Fig. 1A) 
\begin{equation}
\mathbf{f}(\mathbf{x},t)=\lim_{N\to \infty} \sum_j^N (\mathbf{F}_j +\mathbf{N}_j) \delta(\mathbf{x} - \mathbf{r}_j)
\label{eq:force}
\end{equation}
that introduces a jump in the fluid stress field $\boldsymbol{\sigma}= -p \mathbf{I} + \eta (\nabla \mathbf{u} + (\nabla \mathbf{u})^T)$, where $p$ is the pressure field and $\mathbf{I}$ the identity matrix; namely,  
$
\llbracket 
\boldsymbol{\sigma} \cdot \mathbf{e}_z 
\rrbracket 
{}_{z=h} 
= \mathbf{f}(\mathbf{x})$.
We map the planar components of $p$ and $\mathbf{u}$ to a 2D Fourier space $\mathbf{k} = (k_x,k_y)$ and $k = \| \mathbf{k} \|$.
Note that the doubly-periodic domain $L\times L$ in the discrete model affords wavenumbers $k_x, k_y$ in the range $[2\pi/L,\pi/d]$.
We solve the 3D Stokes equation analytically in the fluid regions above and below the force density layer, at which we match the jump and boundary conditions (SI, $\S$1).  
We arrive at $\hat{\mathbf{v}} = \widehat{\mathbf{K}} \cdot \hat{\mathbf{f}}$,
where $\hat{\mathbf{f}}$ and $\hat{\mathbf{v}}$ are the Fourier transform of  $\mathbf{f}$ and $\mathbf{v}$ and $\widehat{\mathbf{K}}$ encodes the Stokes kernel associated with the force density layer, 
\begin{equation}
\begin{split}
\label{eq:K}
    \widehat{\mathbf{K}} & = -\dfrac{e^{-2k h}}{4 \eta k^3} \left[\dfrac{}{} 2k^2 (1 - e^{2k h}) \mathbf{I}  \ +  \right. \\
    & \hspace{0.5in}\left. (-1 + e^{2k h} + 2h k (-1 + h k))\mathbf{k} \otimes \mathbf{k} \dfrac{}{}\right],
\end{split}
\end{equation}
and $ \lim_{\mathbf{k} \to 0} \widehat{\mathbf{K}} = h/\eta \mathbf{I}$; see Figs.~S3-S5 of SI for a direct comparison between the flow fields in the continuum and discrete models.
This analytically-tractable expression for the cilia-generated flow field $\mathbf{v}$, together with~\eqref{eq:K} and the force that cilia exert on the fluid~\eqref{eq:force}, form the basis for deriving a continuum theory for phase coordination in ciliated surfaces in terms of the moments fields $Y_n(\mathbf{x},t)$.

To derive evolution equations for $Y_n(\mathbf{x},t)$, we differentiate~\eqref{eq:Yn} with respect to time, substitute~\eqref{eq:eom_discrete_phi} in the resulting equation, and take the limit of large $N$. We arrive at the continuum model (SI, $\S$4)
\begin{equation}
\begin{split}
\label{eq:Yhatdot}
    \dot{Y}_n &= i n \Omega_o \bigg[ Y_{n}(\mathbf{x})  + \sum_{m} \left(\widetilde{\mathbf{a}}_{m} \cdot \mathbf{v} + \widetilde{\mathbf{B}}_{m}:   \nabla{\mathbf{v}} \right)Y_{n+m} \bigg],
\end{split}
\end{equation}
where the vectors $\widetilde{\mathbf{a}}_{m}$ and  second-rank tensors $\widetilde{\mathbf{B}}_{m}$ are coefficients of the Fourier series expansions of ${\mathbf{t}}/{b \Omega}$ and ${\mathbf{t} \otimes \mathbf{n}}/{\Omega} $  with respect to $\phi$, respectively.
The cilia-generated force density $\mathbf{f}$ in~\eqref{eq:force} can be expressed in terms of $Y_n$ as
\begin{align}
	\mathbf{f} = \rho_c \zeta \sum\limits_{m}  & \left[ \left( \widetilde{\mathbf{c}}_{m} - \widetilde{\mathbf{D}}_{m} \cdot \mathbf{v}(\mathbf{x}) \right) Y_m + \right. \label{eq:fCoarse_final} \\ \nonumber
	& \left. b \nabla \cdot  \left( (\widetilde{\mathbf{E}}_{m} - \widetilde{\mathbf{G}}_{m} \cdot \mathbf{v}(\mathbf{x}) )Y_m \right) \right],  
\end{align}
where $\widetilde{\mathbf{c}}_{m}$, $\widetilde{\mathbf{D}}_{m}$, $\widetilde{\mathbf{E}}_{m}$, $\widetilde{\mathbf{G}}_{m}$ 
are the coefficients of the Fourier series expansions of 
	$b \Omega \mathbf{t}$, 
	$\mathbf{n}\otimes \mathbf{n}$, 
	$b \Omega \mathbf{n} \otimes \mathbf{t}$, and
	$2 \mathbf{n} \otimes \mathbf{n}\otimes \mathbf{n}$, respectively. Explicit expressions are given in SI.

\eqref{eq:Yhatdot}, together with \eqref{eq:K} and \eqref{eq:fCoarse_final},
form a set of linear partial differential equations that describes the dynamical evolution of the moment fields $Y_n(\mathbf{x},t)$. Since all $Y_n$ are coupled, a closure assumption is needed to arrive at a tractable set of equations. When using closure assumption appropriate for locally coordinated phase waves, we recover the theory presented in \cite{Meng2021} as a special case.
Here, we are interested in the stability of the isotropic and synchronized states, respectively, and the emergence of coordination; we will thus make closure assumptions appropriate for these cases.

\paragraph{Stability of isotropic states} The isotropic state is a periodic steady state of~\eqref{eq:Yhatdot}, with $Y_0(\mathbf{x}) = 1$ while
$Y_n$, $\mathbf{f}$, and $\mathbf{v}$ are all identically zero.
We consider small perturbations $\delta \hat{Y}_n$  about the isotropic state, and linearize \eqref{eq:Yhatdot} accordingly (see SI). The resulting linear equation describing the time evolution of $\delta Y_n(\mathbf{x},t)$ is best expressed in 
Fourier space in terms of the associated Fourier transform $\delta\hat{Y}_n(\mathbf{k},t)$.
We arrive at the eigenvalue problem in Fourier space
$
   \delta\dot{\hat{Y}}_n = \sum_{m=-\infty}^{\infty} \hat{L}_{nm}\delta\hat{Y}_m, 
$
where $\hat{L}_{nm}$ depends on $\widehat{\mathbf{K}}$ (see SI).

\begin{figure*}[t]
	\centering
	\includegraphics[scale=1]{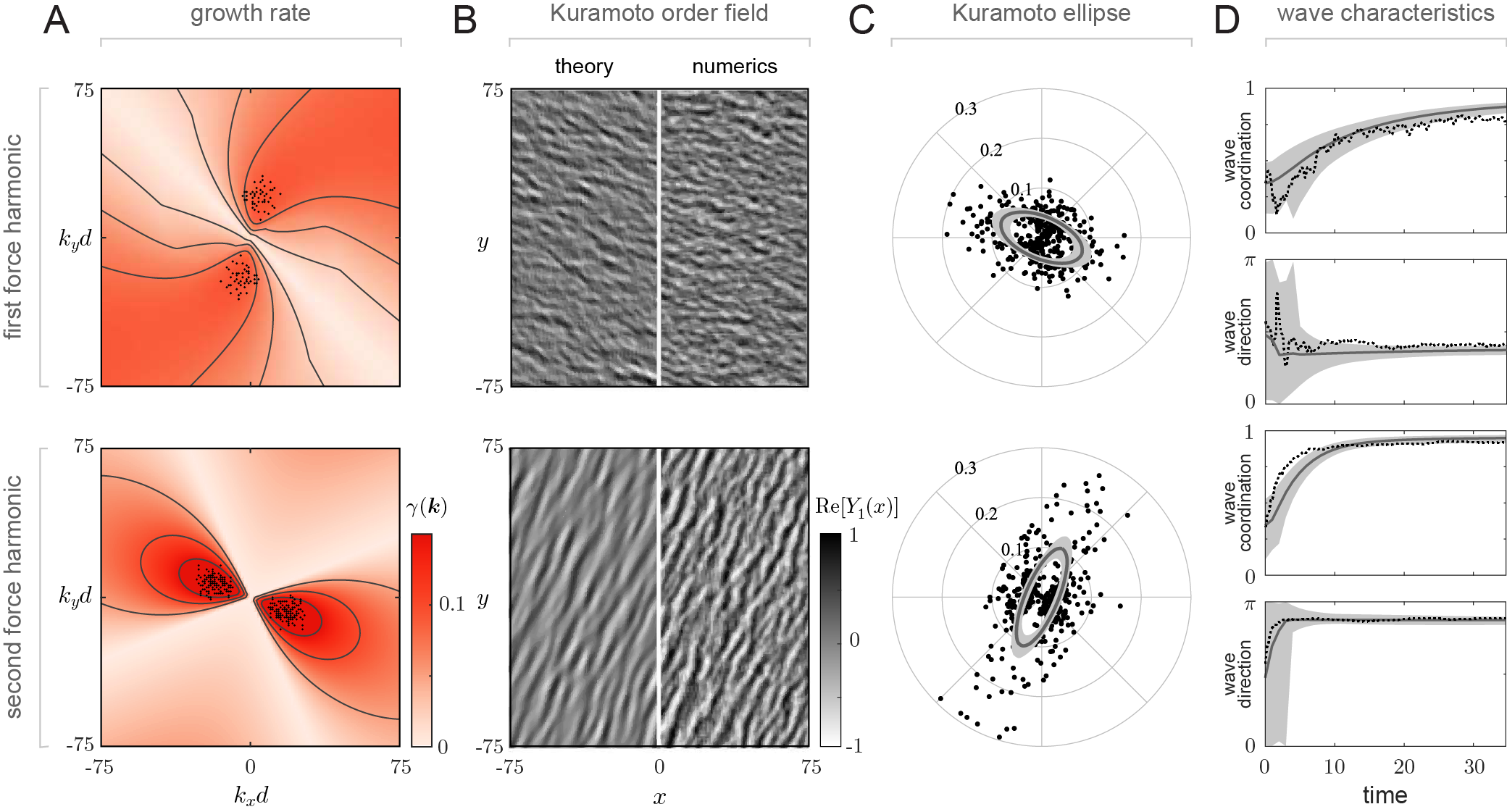}
	\caption{\textbf{Wave instability from isotropic state:}
	\textbf{A.} Growth rate $\gamma(\mathbf{k})$ of the instability predicted by the continuum theory (colormap) and discrete simulations (black dots) in the Fourier space.
	\textbf{B.} Starting from a random Kuramoto field $Y_1(\mathbf{x})$, map to Fourier space $\hat{Y}_1(\mathbf{k})$ and use $\gamma(\mathbf{k})$ from panel A to step forward in time. Snapshots of Re$|Y_1(\mathbf{x},t)|$ show the emergence of traveling waves similar to those obtained from direct numerical simulations of \eqref{eq:eom_theta}.
	\textbf{C.} Kuramoto ellipse introduced in Fig.~1E  obtained from the sample simulation in panel B (black dots) and theory  corresponding to 200 realizations (grey), with average shown in dark grey.
	\textbf{D.} Time evolution of wave coordination and the angle of the Kuramoto ellipse; sample simulation (black line) and theory (grey). } \label{fig:isotropic}
\end{figure*}

Near the isotropic state, $\hat{Y}_n$ decays fast with $n$. We thus close the  linearized system of equations by truncation. We solve the truncated equations numerically over the entire Fourier space.
In Fig.~\ref{fig:isotropic}A, we report the growth rate $\gamma(\mathbf{k})$ (maximal real part of the eigenvalues) as a colormap over all wavenumbers $\mathbf{k}$ for the first and second force harmonics in Fig.~\ref{fig:setup}.  
The isotropic state is unstable for almost all wavenumbers, with stronger instability exhibited by the second force harmonic. 
We superimpose data from a sample discrete cilia simulation (black dots) of~\eqref{eq:eom_theta}; the dots represent the wavenumbers at which $|\hat{Y}_1(\mathbf{k},t)|$ is maximal (exceeding a given threshold). Note the agreement between the simulation result and the maximal growth rate in the continuum theory.

\paragraph{Linear instability governs emergent  coordination} To further challenge the theory, starting from the initial state used in the sample simulation, we calculate $\hat{Y}_1(\mathbf{k},0)$, and integrate forward in time 
$\hat{Y}_1(\mathbf{k},t) = \hat{Y}_1(\mathbf{k},0) \textrm{exp}({\gamma(\mathbf{k}) t})$ according to the linear growth rate from the continuum model.
We transform back to physical space and compare $\text{Real}[Y_1(\mathbf{x},t)]$ to that obtained from the sample simulation (Fig.~\ref{fig:isotropic}B) and again find remarkable agreement, indicating that the wave patterns are governed by the fastest growing linear mode.
Lastly, we calculate the Kuramoto ellipse based on Monte-Carlo computations using the continuum model growth rate with 200 initial isotropic states; the average is shown in dark grey in Fig.~\ref{fig:isotropic}C and the range in lighter grey while results from the sample simulation are superimposed in black dots.
The time evolution of the ellipses' eccentricity and angle are shown in  Fig.~\ref{fig:isotropic}D using the same color convention.
The ellipse eccentricity correlates with the instability growth rate and its angle in physical space is orthogonal to the direction of maximum growth rate in Fourier space.
We conclude that the fastest growing wavenumbers in the linear continuum theory are predictive of the long-term traveling wave patterns in the nonlinear discrete model.

\begin{figure}[t]
 	\centering
 	\includegraphics[width=\linewidth]{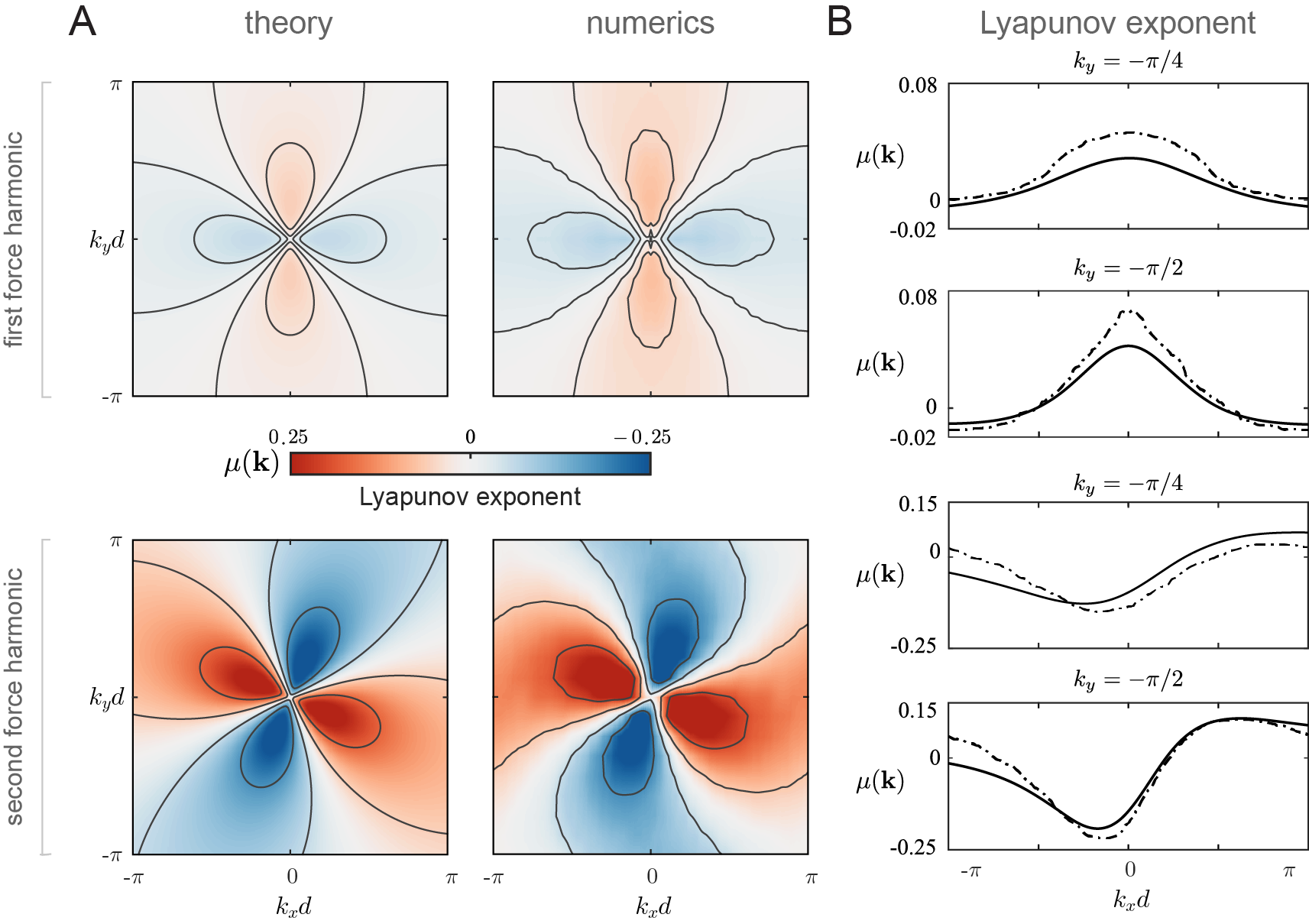}
 	\caption{\textbf{Wave instability from syncrhonized state:} Lyapunov exponent $\mu(\mathbf{k})$  obtained from theory and simulations \textbf{A.} over the entire wave space (colormap), and 
 	\textbf{B.} at cross-sections $k_y d = -\pi/4$ and $k_y = -\pi/2$ with theory in solid lines and simulations in dashed lines.
 }
 	\label{fig:sync}
 \end{figure}

\paragraph{Stability of synchronized states} 
For completeness, we carry out a rigorous stability analysis of the synchronized state. This state could be analyzed using the techniques provided in~\cite{Meng2021}, but here we provide an alternative approach consistent with our continuum theory (SI, $\S5$).
Synchronous states are unstable to general perturbations, with stronger instability in the case of the second force harmonic, consistent with the isotropic state. 
The growth rates $\mu(\mathbf{k})$
are shown in Fig.~\ref{fig:sync}A over the entire wave space and in Fig.~\ref{fig:sync}B for select cross-sections.

\begin{figure*}[!ht]
	\centering
	\includegraphics[scale=1]{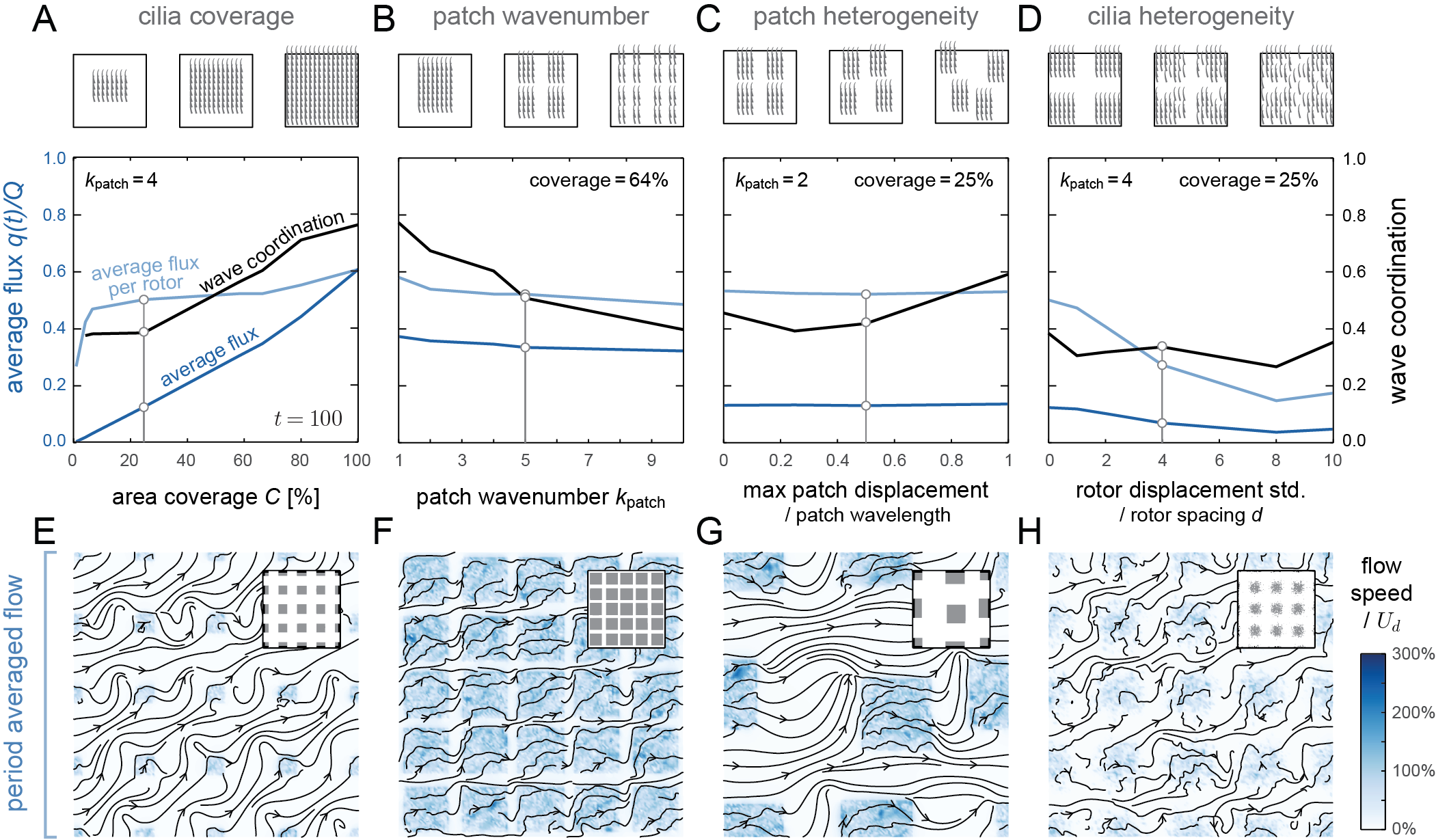}
	\caption{\textbf{Effects of tissue heterogeneity on synchronization and pumping.}
	\textbf{A.} The area fraction of the periodic domain that is covered by cilia strongly affects the pumping performance and wave coordination (black line) of the ciliary carpet. The net average flux as measured by total flux over the all possible 151$\times$151 possible sites for rotors are roughly linearly correlated with the area coverage percentage (dark blue line). However, if we normalize net flux against the number of active rotors present (flux per rotor, light blue line), there is no significant decrease of flux even when area coverage fraction drops to near 20\%. This shows that fluid pumping are equally efficient as long as enough number of rotors are inside each patch for metachronal waves to form.
	\textbf{B.} As the number of patches increase, both net flux and wave coordination see a slight decrease. 
	\textbf{C.} When patches are distributed in a heterogeneous fashion, the wave coordination improves without a strong impact on the net flux. 
    Here the $x-$axis is the maximum possible patch displacement normalized by the patch wavelength to avoid merged patches.
	\textbf{D.} When individual rotors are displaced randomly by Gaussian noise, the average flux decreases while the wave coordination measure remains roughly unchanged. Here the $x-$axis is the rotor displacement standard deviation normalized by the inter-rotor spacing $d$.
	\textbf{E-H.} Period averaged flow streamlines shown for the marked cases in A-D.
	All panels shown use first force harmonic only.
	} \label{fig:patch}
\end{figure*}

\paragraph{Effect of tissue heterogeneity on cilia coordination and fluid pumping} 
=Lastly, we examine the effect of heterogeneity in the geometric distribution of cilia~\cite{Ramirez2020,Boselli2021} on the emergent coordination and fluid pumping. We study tissue heterogeneity in the context of the discrete model~\eqref{eq:eom_theta}. 
Inspired by the empirical observations of~\cite{Ramirez2020}, we introduce four geometric parameters that allow us to represent and manipulate the distribution of ciliary patches and individual cilia (SI, $\S$6). First, we define a cilia coverage ratio $C$ that determines the area fraction that is covered by active cilia (Fig.~\ref{fig:patch}A). We distribute the cilia into regular patches of wavenumber $k_{\textrm{patch}}$ in the $x$- and $y$-directions (Fig.~\ref{fig:patch}B). We randomly perturb the location of each entire patch in the $x$- and $y$-directions (Fig.~\ref{fig:patch}C), and we introduce Gaussian noise in the location of individual cilia (Fig.~\ref{fig:patch}D).

A systematic investigation of how wave coordination and fluid pumping vary with these geometric heterogeneity parameters is presented in Fig.~\ref{fig:patch}A-D, and flow streamlines and speed for select parameter values are shown in Fig.~\ref{fig:patch}E-H.
In all cases, rotor height $h$ and distance between each rotor $d$ were kept constant, and rotors were driven using the first force harmonic $F_1 = 1 + 0.5\cos\theta$.
Fig.~\ref{fig:patch}A shows that wave coordination and net flux scale roughly linearly with coverage ratio $C$, but renormalizing by the number of active rotors reveals that the net flux per rotor is almost independent of $C$ even for area coverage as low as 20\%. 
At $C=25\%$, with cilia distributed in four ciliary patches per direction, the flux per rotor is more than 80\% of that of the fully covered carpet (Fig.~\ref{fig:patch}E). 
This suggests that fluid pumping may be more efficient under partial coverage of ciliated tissues, as long as cilia are organized into relatively-large and coherent patches. 
For a given value of coverage ratio, both net flux and flux per rotor are affected little by distributing the cilia into regularly- and irregularly-spaced patches (Fig.~\ref{fig:patch}B,C). Wave coordination is slightly reduced with increased number of patches (Fig.~\ref{fig:patch}B,F), but enhanced with increased irregularity in patch spacing (Fig.~\ref{fig:patch}C,G).
A closer look at the streamlines shows eddies at the boundaries of small ciliary patches (Fig.~\ref{fig:patch}E, H). These eddies hinder coherent fluid pumping, but are minimized when cilia are distributed in larger ciliary patches (Fig.~\ref{fig:patch}F,G) even when the coverage ratio is significantly compromised (Fig.~\ref{fig:patch}G). 
Importantly, while irregularity in the spacing of ciliary patches seems beneficial, a noisy distribution of individual cilia significantly hinders fluid pumping (Fig.~\ref{fig:patch}D,H). 
These findings suggest that wave coordination and fluid pumping remain robust when cilia are distributed into separate ciliated patches that consist of large clusters of regularly-spaced cilia within each patch, even when the patches themselves are not regular. This is exactly the design adopted in healthy ciliated tissues: ciliated cells cluster in irregular patches of regularly-space individual cilia, separated by other cell types~\cite{Ramirez2020,Nawroth2020}.

\section*{Conclusions}

We analyzed phase coordination and fluid pumping in large scale ciliary carpets in the context of a discrete rotor model and a novel continuum theory.
Existing models have been limited to a few hundred cilia~\cite{Elgeti2013,Osterman2011}, which is several orders of magnitude smaller compared to the number of cilia found, for instance, in mammalian airways \cite{Ramirez2020}.
Our models serve as a first step towards establishing maps from single-cilium kinematics to tissue-level phase coordination and fluid pumping in large scale ciliary systems.

To isolate the effect of hydrodynamic coupling on the emergence of collective ciliary states that pump fluids,  we accounted for the single cilium beat kinematics using a rotor model whose phase oscillations do not pump fluid.
This idealization captures important features of actual cilia beat patterns: the first harmonic forcing that drives phase oscillations introduces force polarity and captures asymmetry in ciliary beating, while the second force harmonic is “nematic” and captures the non-planar ellipticity of ciliary beating. This interpretation is important because it has practical implications on the pumping function of the ciliated tissue.

We showed in both numerics and theory that, in ciliary carpets, spatially-homogeneous isotropic and synchronized states are unstable.  By construction, these states do not pump fluids. However, the instabilities that arise due to hydrodynamic interactions lead to the emergence of wave patterns that can break time reversal symmetry and pump flows.
Our theory in the limit of infinite number of cilia predicts quantitatively the long-term traveling wave patterns observed in the nonlinear numerical simulations, indicating that linear instabilities dominate the emergent waves. 

The emergent traveling wave patterns depend on the properties of the single cilium but are robust to large perturbations in initial conditions. 
That is, in ciliary carpets, a robust traveling wave pattern exists as a stable global attractor, as reported in~\cite{Solovev2020}. 
The specific properties of this attracting state depend on the characteristics of the single cilium: ciliary beat asymmetry (first force harmonic) produces waves that propagate in the same direction and pump fluids, while ciliary beat ellipticity (second force harmonic) produces waves that propagate in opposite directions and do not pump fluid.

Compared to the theory presented in~\cite{Meng2021}, our work provides a rigorous formulation of the fluid velocity field induced by a continuum ciliary force layer, and a mechanism for accounting for the onset of instabilities from non-coordinated, isotropic states. 
Importantly, the instability we detected at the first force harmonic is not reported in~\cite{Meng2021}.  This instability, though subsiding the instability at the second force harmonic, governs the direction of fluid pumping and is therefore essential for tissue-level function. Our findings emphasize the importance of considering the effect of single-cilium beat kinematics on the coupling between wave coordination and fluid pumping in ciliated tissues.

We introduced tissue-level heterogeneity based on empirical observations of mammalian airway tissue~\cite{Ramirez2020}. We found that the emergent wave coordination and fluid pumping are robust to tissue heterogeneity. These findings support the notion that the geometric design of ciliated tissues, where ciliated cells cluster in irregular patches of regularly-space individual cilia separated by other cell types~\cite{Ramirez2020,Nawroth2020}, is robust in terms of wave coordination and fluid pumping. However, designs not seen in healthy tissues where individual cilia are irregularly distributed are inferior in both.

Taken together, the dependence of emergent states on the properties of the single cilium and the robustness of these states to large perturbations and tissue heterogeneity suggest that functional specificity of the ciliated tissue is interlaced with the beating kinematics of the single cilium and the organization of the ciliated tissue. They corroborate existing assays that examine the beat kinematics of individual cilia to distinguish between diseased and healthy states~\cite{Chioccioli2019,Chioccioli2019b,Blanchon2020}, and suggest to amend these assays with new measures of tissue-level organization.

Our results that a specific wave state emerges as a global attractor is crucial for specialized fluid pumping in ciliated tissues, but multiple coordination states are essential for organisms that use cilia for locomotion.
Multiple coordination states are typically found in models of few filaments and rowers that beat in a plane perpendicular to the tissue surface~\cite{Guo2018,Guo2020,Solovev2020, Man2020,Chakrabarti2021}. 
In addition to the mere size of the ciliary system, the cilium beat kinematics is a marked difference between these studies and ours. In our model, each rotor beats cyclically in a plane parallel to the tissue surface, akin to cilia with non-planar beating kinematics, and the strongest instability is driven by this non-planar property (accounted for by the second force harmonic). These observations
indicate that cilia function -- fluid pumping in ciliated tissues  versus locomotion in multi-ciliated organisms  – is interwoven in the architecture of the ciliated system. They suggest the importance of future studies that systematically probe the effects of the size of the ciliary carpet and planarity of the cilium beating kinematics on the uniqueness of the emergent coordination state.

We made a few simplifications that can be readily relaxed in future extensions of this study. For example, 
our theory can be readily extended to incorporate non-circular limit-cycle representations of individual cilia obtained from experimental data of cilia beating patterns~\cite{Brumley2014} and non-uniform lattice structures reconstructed from high-resolution images of ciliated tissues~\cite{Ramirez2020}. Specifically, cilium kinematics with asymmetric beating patterns that pump fluid can be readily incorporated. Our tissue heterogeneity analysis can be extended 
to test the effect of variations in cilia alignment and frequency~\cite{Loiseau2020,Ramirez2020} on wave coordination and fluid pumping.
These extensions will allow us to quantitatively calibrate the model for specific ciliated tissues, and to predict how perturbations to the single-cilium kinematics, e.g.,~\cite{Solovev2020, Solovev2022}, and tissue organization, e.g.,~\cite{Boselli2021}, affect the emergence of large scale coordination and fluid pumping.

{\bf Acknowledgment} E.K. acknowledges support from the National Science Foundation (INSPIRE award MCB-1608744, RAISE award IOS-2034043, and CBET-2100209), the National Institutes of Health (R01 grant HL 153622-01A1), the Office of Naval Research (grant 12707602 and grant N00014-17-1-2062), and the Army Research Office (ARO W911NF-16-1-0074).
S.F. is supported by the Vienna Science and Technology Fund (WWTF) and the City of Vienna through project VRG20-002.1
We thank Janna A. Nawroth, Michael J. Shelley, and Jingyi Liu for helpful discussions. 

\bibliography{references}

\end{document}